# Middleware-based Database Replication: The Gaps Between Theory and Practice

Emmanuel Cecchet

EPFL
Lausanne, Switzerland

`emmanuel.cecchet@epfl.ch`

George Candea

EPFL & Aster Data Systems
Lausanne, Switzerland

`george.candea@epfl.ch`

Anastasia Ailamaki

EPFL & Carnegie Mellon University
Lausanne, Switzerland

`anastasia.ailamaki@epfl.ch`

## ABSTRACT

The need for high availability and performance in data management systems has been fueling a long running interest in database replication from both academia and industry. However, academic groups often attack replication problems in isolation, overlooking the need for completeness in their solutions, while commercial teams take a holistic approach that often misses opportunities for fundamental innovation. This has created over time a gap between academic research and industrial practice.

This paper aims to characterize the gap along three axes: performance, availability, and administration. We build on our own experience developing and deploying replication systems in commercial and academic settings, as well as on a large body of prior related work. We sift through representative examples from the last decade of open-source, academic, and commercial database replication systems and combine this material with case studies from real systems deployed at Fortune 500 customers. We propose two agendas, one for academic research and one for industrial R&D, which we believe can bridge the gap within 5-10 years. This way, we hope to both motivate and help researchers in making the theory and practice of middleware-based database replication more relevant to each other.

### Categories and Subject Descriptors
C.2.4 [Distributed Systems]: Distributed databases; H.2.4 [Systems]: Distributed databases

### General Terms
Performance, Design, Reliability.

### Keywords
Middleware, database replication, practice and experience.

## 1. INTRODUCTION

Despite Gray's warning on the dangers of replication [18] over a decade ago, industry and academia have continued building replication systems for databases. The reason is simply that replication is the only tried-and-true mechanism for scaling performance and availability of databases across a wide range of requirements.

There exist replication "solutions" for every major DBMS, from Oracle RAC™, Streams™ and DataGuard™ to Slony-I for Postgres, MySQL replication and cluster, and everything in-between. The naïve observer may conclude that such variety of replication systems indicates a solved problem; the reality, however, is the exact opposite. Replication still falls short of customer expectations, which explains the continued interest in developing new approaches, resulting in a dazzling variety of offerings.

Even the "simple" cases are challenging at large scale. We deployed a replication system for a large travel ticket brokering system at a Fortune-500 company faced with a workload where 95% of transactions were read-only. Still, the 5% write workload resulted in thousands of update requests per second, which implied that a system using 2-phase-commit, or any other form of synchronous replication, would fail to meet customer performance requirements (thus confirming Gray's prediction [18]). This tradeoff between availability and performance has long been a hurdle to developing efficient replication techniques.

In practice, the performance/availability tradeoff can be highly discontinuous. In the same ticket broker system mentioned above, the difference between a 30-second and a one-minute outage determines whether travel agents retry their requests or decide to switch to another broker for the rest of the day ("the competition is one click away"). Compounded across the hundreds of travel agencies that connect to the broker system daily for hotel bookings, airline tickets, car rentals, etc., the impact of one minute of downtime comes close to that of a day-long outage. The replication system needs to be mindful of the implied failover requirements, and obtaining predictable behavior is no mean feat.

Our premise is that, by carefully observing real users' needs and transforming them into research goals, the community can bridge the mismatch between existing replication systems and customers' expectations within the coming decade. We sift through the last decade of database replication in academic, industrial, and open-source projects. Combining this analysis with 45 person-years of experience building and deploying replicated database systems, we identify the unanswered challenges of practical replication. We find that a few "hot topics" (e.g., reliable multicast and lazy replication [21]) attract the lion's share of academic interest, while other equally important aspects (e.g., availability and management) are often forgotten—this limits the impact research systems can have on the real world. Motivated by these findings, we draft possible agendas for academic and industrial research.

This paper concentrates exclusively on middleware-based[1] replication for OLTP workloads. The prevalent architecture is shared-nothing, where cluster nodes use local disks to store data. We make two contributions. First, we identify gaps between database research and practice on four different levels: RDBMS engine, SQL language, middleware, and system management. We show how overlooking seemingly small details can undermine replication systems. Second, we distill a few research topics that provide low-hanging fruit for closing these gaps, in the realms of middleware design, consistency models, availability, and system evaluation. We also describe what we believe industry ought to do with respect to interfaces, transaction abstractions, system management, and dynamic upgrades. As this paper is not intended to serve as an area survey, the reference list is by no means exhaustive. Rather, we choose representative examples that help us characterize the academia/industry gap and its consequences.

The rest of the paper is structured as follows: Section 2 describes the replication schemes presently favored in the field. Section 3 surveys representative academic proposals for replication. Section 4 discusses in detail the practical challenges we have encountered while deploying numerous middleware-based replication systems at customers ranging from small startups to Fortune 500 companies. Section 5 distills the main challenges and outlines roadmaps for both academic research and industrial R&D that can bridge the identified gaps. Section 6 concludes the paper.

## 2. REPLICATION IN PRACTICE

There are two main reasons to employ database replication: to improve performance and to increase availability. Section 2.1 discusses commonly used architectures for performance-focused deployments, while Section 2.2 describes availability-focused solutions.

## 2.1 Improving Performance via Replication

Database replication is typically used to improve either read performance or write performance, while improving both read and write performance simultaneously is a more challenging task.

Figure 1 depicts *master-slave replication*, a popular technique used to improve read performance. In this scenario, read-only content is accessed on the slave nodes and updates are sent to the master. If the application can tolerate loose consistency, any data can be read at any time from the slaves given a freshness guarantee. As long as the master node can handle all updates, the system can scale linearly by merely adding more slave nodes. Examples of commercial products providing asynchronous master-slave replication are Microsoft SQL Server replication, Oracle Streams, Sybase Replication Server, MySQL replication, IBM DB2 DataPropagator, GoldenGate TDM platform, and Veritas Volume Replicator.

A special instance of read throughput improvement relates to legacy databases: often an old DB system is faced with increased read performance requirements, that it can no longer satisfy, yet replacing the DB is too costly. Recently emerged strategies, such as satellite databases [29], offer a migration path for such cases. In the case of an e-commerce application, the main legacy database is preserved for all critical operations, such as orders, but less critical interactions, such as catalog browsing, can be offloaded to replicas. Such configurations typically use partial replication—all orders could be solely on the main legacy database, while only the catalog content is replicated. As an application might also be using multiple database instances inside the same RDBMS, the user can choose to replicate only specific database instances.

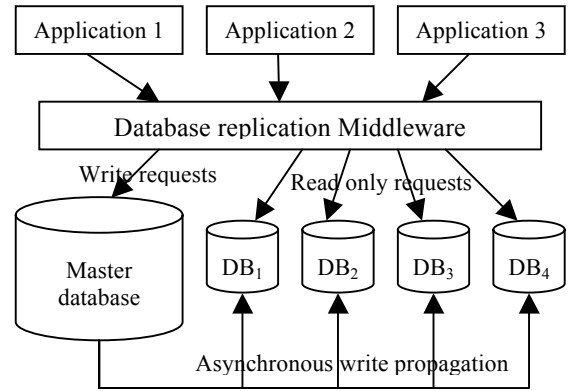

**Figure 1. Database scale-out scenario**

*Multi-master replication* allows each replica owning a full copy of the database to serve both read and write requests. The replicated system then behaves as a centralized database, which theoretically does not require any application modifications. Replicas, however, need to synchronize in order to agree on a serializable execution order of transactions, so that each replica executes the update transactions in the same order. Also, concurrent transactions might conflict, leading to aborts and limiting the system's scalability [18]. Even though real applications generally avoid conflicting transactions, there are still significant research efforts trying to solve this problem in the replication middleware layer. The volume of update transactions, however, remains the limiting performance factor for such systems. As every replica has to perform all updates, there is a point beyond which adding more replicas does not increase throughput, because every replica is saturated applying updates.

Examples of commercial multi-master architectures include Continuent uni/Cluster and Xkoto Gridscale for middleware replication, and MySQL Cluster and DB2 Integrated Cluster for database in-core implementations. Shared-disk architectures, such as Oracle RAC, are out of the scope of this paper.

Finally, *data partitioning* techniques can be used to address write scalability. Figure 2 shows an example where data is logically split into 3 different partitions, each one being replicated. Common partitioning criteria are based on a table primary key and include techniques such as range partitioning, list partitioning and hash partitioning. The benefits of this approach are similar to RAID-0 for disks: updates can be done in parallel to partitioned data segments. Read latency can also be improved by exploiting intra-query parallelism and executing the sub-queries in parallel on each partition.

---

[1] By middleware we mean the software layer that lies between an application and the database replicas.

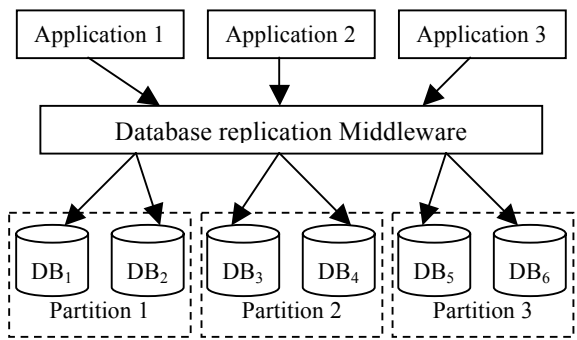

Figure 2. Database partitioning for increased write performance

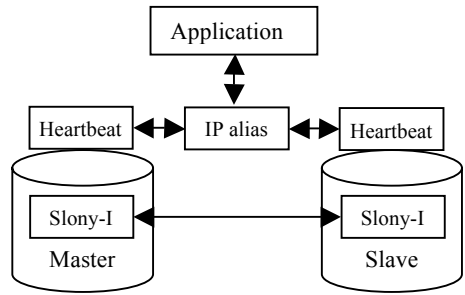

Figure 3. Hot standby configuration

## 2.2 Increasing Availability via Replication

High availability requires low downtime. Planned downtime is incurred during all software and hardware maintenance operations, while unplanned downtime can strike at any time and is due to predictable or unpredictable failures (hardware failures, software bugs, human error, etc.). Downtime is usually the primary metric being optimized for, with performance relegated to a secondary role (although, in practice, both are desired).

A system's availability is the ratio of its uptime to total time. In practice, it is computed as the ratio between the expected time of continuous operation between failures to total time, or

$$Availability = \frac{MTTF}{MTTF+MTTR} \Rightarrow Unavailability = \frac{MTTR}{MTTF+MTTR} \approx \frac{MTTR}{MTTF}$$

where MTTF is mean-time-to-failure and MTTR is mean-time-to-repair. Since MTTF >> MTTR, the unavailability (ratio of downtime to total time) is approximately MTTR/MTTF.

The goal of replication, together with failover and failback, is to reduce MTTR and thus reduce unavailability. Failover is the capability of switching users of a database node over to another database node containing a replica of the data, whenever the node they were connected to fails. Failback is the reverse and occurs when the original replica recovered from its failure and users are re-allocated to it.

*Hot standby* is the most commonly deployed configuration using database replication in both open-source and commercial databases. Whether it uses single-master or multi-master techniques, the end goal remains the same: to provide fast recovery from node failures. A node serves all queries and, upon failure, the workload is transferred to the hot standby node.

Figure 3 shows a hot standby setup using Slony-I for PostgreSQL [31]. There are two replicas, one acting as a master and the other as a slave. The application connects to a simple load balancer that directs the requests to the master and, when the master's failure is detected, requests are rerouted to the slave.

The hot standby or slave node, either local or at a remote site, has computing capabilities similar to the master node. It applies updates as they arrive from the master. As soon as the master fails (detected by a simple heartbeat mechanism), the load is shifted to the slave node. Various techniques can be used for this, such as virtual IP [10] or reconfiguration of the driver or application.

Determining which transactions are lost when the master fails remains a manual procedure that requires careful inspection of the master's transaction log (when it is available). The best guarantee that is usually offered is 1-safe (i.e., transactions commit at the master without consulting the slave) with an upper-bound time window (e.g., at most all transactions committed in the past 5 minutes have been lost). These guarantees are usually considered weak, but good enough for maintaining uptime in the face of single faults. 2-safe database replication forces the master to commit only when the backup has also confirmed receipt of the update (even though the backup may not have written to disk immediately). This avoids transaction loss, but increases latency.

In order to not completely waste the slave's computing resources under normal operation, the slave is typically used for read-only reporting tasks that run in batch mode. In a heavily loaded production system, however, the lag between the master and slave node can become significant. Our customers report several hours of downtime when commercial databases failover clients from a master to a hot-standby slave node. The reason is typically that the trailing updates are applied serially at the slave, whereas the master processes them in parallel. The "solution" is usually to slow down the master (during normal operation) so as to keep the slave synchronized to within a small time window.

*WAN replication* is the gold standard for sustaining availability in the face of disasters that would affect an entire cluster (earthquakes, floods, etc.). In the case of disaster recovery, unlike regular failover, clients' requirements for synchronization windows are less stringent. Replicating data asynchronously between sites, possibly by interconnecting middleware replication solutions, usually involves both data partitioning and multi-way master/slave replication (i.e., each site is master for its local geographical data)—see Figure 4.

Major challenges in WAN replication are partitions resulting from network failure or datacenter infrastructure failures. Although some companies can afford their own data centers interconnected by dedicated leased lines, most use professional hosting centers, to reduce costs. Hosting centers emphasize efficient multiplexing and management of resources so, even if individual applications are not demanding in terms of resources, there are often hundreds or thousands of them running on various combinations of software stacks in one datacenter. In such dense environments, one problem can have domino effects. Our experience with an academic GRID (600 CPUs) and a number of smaller clusters (10-100 nodes) indicates that, on average, one fatal failure (software or hardware) occurs per day per 200 processors, not including air conditioning or power outages. Thus, keeping replicas in sync can be challenging when failures are frequent.

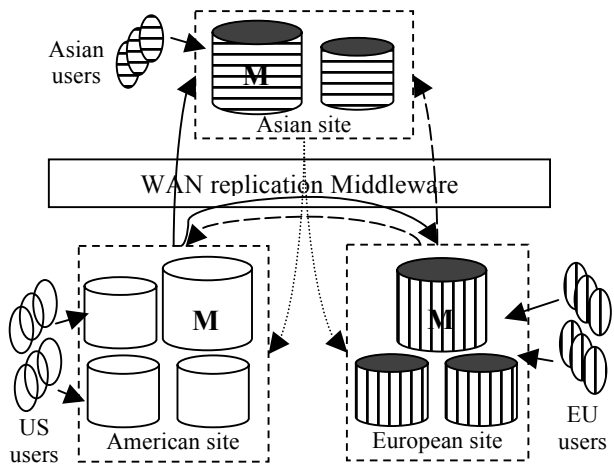

**Figure 4. Worldwide multi-way master/slave replication**

## 3. ACADEMIC APPROACHES TO MIDDLEWARE-BASED REPLICATION

In this section, we describe some of the leading architectures adopted by modern academic prototypes, as well as the major points of research interest. Many academic proposals have not (yet) made their way into practice because of practical challenges we describe later on (Section 4).

### 3.1 System Design

Most of middleware-based replication research focuses on multi-master replication. Master-slave implementations are either in-core implementations or third-party tools that extract database binlogs and ship them to another database. Slony-I [31] and Ganymed [28] for PostgreSQL are among the few middleware solutions that provide direct master-slave replication capabilities. With suitable load balancers, multi-master systems like C-JDBC [8] or Tashkent [14] can also be used in master-slave mode.

Postgres-R [20] was the first significant academic research prototype to provide a fully functional database replication mechanism for an open source database. Even though not implemented 100% in middleware, the replication mechanism interacts with mostly unmodified database core components (subsequently, Middle-R [27] implemented Postgres-R concepts in middleware). As illustrated in Figure 5, the database client/server communication is untouched and the replication only happens behind the scenes, by coordinating the different database engines.

The major advantage of this approach is that it does not require any change on the client side; it does require, however, integration with the database engine. This restricts the ability of different database engines (or even different versions of the same engine) to interact with each other. For this approach to be viable in practice, the replication APIs must be adopted and integrated in the main development line of the database engine itself. In the case of Postgres-R, the failure to transfer the complex piece of replication code to the PostgreSQL core team led to a gradual divergence, eventually rendering Postgres-R obsolete. In the case of closed source databases, this is an even greater challenge.

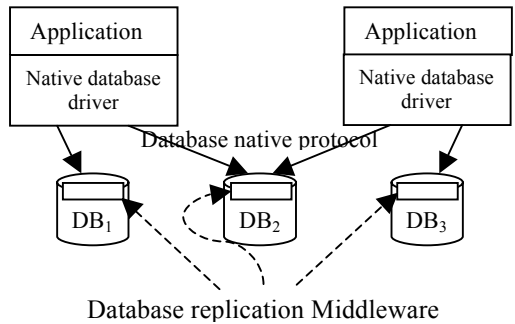

**Figure 5. Query interception at the database engine level**

Figure 6 shows another approach, that intercepts queries directly at the database native protocol level; a prototype intercepting the PHP/MySQL protocol to route queries is described in [3]. Proxying queries at the DBMS native protocol level is elegant, because the middleware is not coupled to the database system and can evolve independently. This approach, however, does not work if the protocol is protected by copyright, licensing or patent restrictions. It also does not support more than one DB engine at the low level.

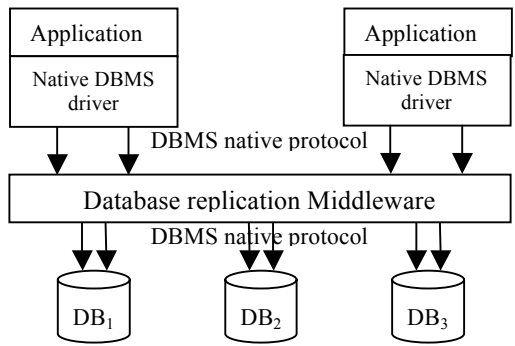

**Figure 6. Query interception at the DBMS protocol level**

It is also possible to intercept the native protocol on the client side, to reuse existing native drivers and remap the calls to a standard API, such as JDBC or ODBC. Myosotis [26], for instance, intercepts MySQL and PostgreSQL protocols and remaps them to the Sequoia/JDBC protocol [30]. This enables the use of Sequoia-specific drivers on supported platforms and the use of native libraries or drivers for other platforms or for accessing non-clustered databases.

Most contemporary academic prototypes nowadays are based on the JDBC proxying concept introduced by C-JDBC [8], depicted in Figure 7. This approach typically requires the database driver to be replaced in the client application. This should normally not require any application code changes, since the new driver implements the same interface as the old one (JDBC, ODBC, etc.). In addition, this approach allows the replication system to span heterogeneous database systems. Some examples of systems designed in this fashion are Tashkent [14], Ganymed [28], and Middle-R [27].

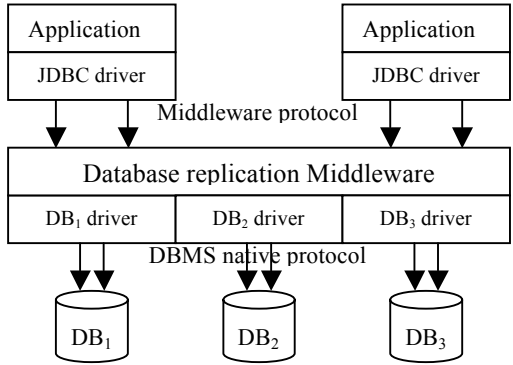

**Figure 7. Query interception in JDBC-based replication**

## 3.2 Load Balancing

A replicated database built for high availability must eliminate all single points of failure (SPOF). Often, projects focusing on performance overlook the replication needs of core components, such as load balancers or certifiers. Even though such redundancy is technically feasible, achieving it is more than mere engineering, because it affects the replication middleware substantially.

To our knowledge, there is no off-the-shelf load balancer for databases, mostly because load balancing is intrinsically tied to the replication middleware. Load balancing can be implemented at the connection level, transaction level or query level. Connection-level load balancing allocates new client connections to replicas according to a specified policy; all transactions and requests on that connection go to the same replica until the connection is closed. This approach is simple, but offers poor balancing when clients use connection pools or persistent connections. Transaction-level or query-level load balancing perform finer grain load balancing by directing queries on a transaction or query basis, respectively.

As an example, Tashkent+ [13] provides transaction-level load balancing and exploits knowledge of the working sets of transactions to allow in-main-memory execution at every replica. The result is an improved throughput of more than 50% over previous techniques; however, the approach uses a centralized load balancer that is not replicated. A failure of this component brings down the entire system. The recovery procedure requires retrieving state from every replica to rebuild the load balancer's soft state. Similar issues would be observed for a certifier failure.

It is possible to achieve near-optimal load balancing with a stateful, centralized load balancer. A failure of the load balancer, however, not only causes all in-flight transactions to be lost, but also causes a complete system outage. Replicating a stateful load balancer or certifier requires extra communication and synchronization that significantly impacts performance. Unfortunately, recovery procedures are rarely described and almost never evaluated in terms of the overhead they introduce in system performance and recovery time.

## 3.3 Data Consistency

Much of today's research chooses snapshot isolation (SI) for enforcing consistency in the database. SI, introduced by [6], is a weaker transactional guarantee than one-copy serializability (1SR), the original standard correctness criterion for replicated data. SI does not offer serializability, but decouples reads from updates to increase concurrency. Each transaction operates on its own copy of data (a *snapshot*), allowing read-only transactions to complete without blocking.

Postgres-R [20] originally proposed an eager replication protocol equivalent to "strong" SI. Various proposed protocols, such as DISCOR and NODO, aim at optimizing performance in this context [19] and are also implemented in Middle-R [27]. [22] extends that work and provides 1-copy SI, also called global strong SI. Ganymed [28] also provides a form of global strong SI called "replicated snapshot isolation with primary copy" (RSI-PC), but it focuses on master/slave architectures and satellite databases. Tashkent [14] relies on generalized snapshot isolation (GSI) and implements prefix-consistent SI (PCSI). [11] proposes global weak SI and evaluates it using simulation. C-JDBC [8] provides pluggable consistency protocols and uses 1SR by default.

## 3.4 Prototype Evaluation

Evaluation of research prototypes mostly relies on benchmarks from the Transaction Processing Council [33]. TPC-W or RUBiS [2] are used for web-related workloads (see Tashkent [14], Ganymed [28], C-JDBC [8]). Micro-benchmarks are also widely used to measure replicated system performance (see Postgres-R [20], Middle-R [27]). System performance is evaluated in terms of throughput (transactions per second, web interactions per seconds, etc.) and latency. If the system under test forms a closed-loop system with the load generator(s), then latency can be directly inferred from throughput.

Scalability measurements almost always use a scaled load to find the best achievable performance (e.g., 5 times more requests for a system with 5 replicas). This usually hides the system overhead at low or constant load. As most production systems operate at less than 50% load, it would be interesting to know how the proposed prototypes perform when under-loaded. To the best of our knowledge, management operations such as backup/restore or adding a node to the system are practically never measured either.

Availability aspects of replication are usually not evaluated in academic prototypes; even recent papers on adaptive middleware [25] focus on performance adaptation in case of workload variations, but do not address adaptation in the presence of failures. In fact, important parameters for evaluating database replication systems (such as mean-time-between-failure, mean-time-between-system-abort, or mean-time-between-critical-failure) are not used, despite them being well explained in the literature. MTTR and MTTF would also seem to be natural metrics for the evaluation of recovery effectiveness. We propose further options in Section 5.1

## 4. PRACTICAL CHALLENGES

Why, given so many well-explained, thoroughly evaluated academic proposals, is database replication still such a challenge in practice? In this section, we use a bottom-up approach to describe the various challenges we have encountered in the field that we believe constitute the primary hurdles in bringing sophisticated replication to real systems. Figure 8 shows the main domains we have identified.

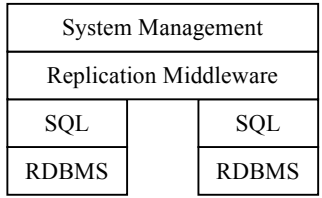

**Figure 8. Layering of practical challenges**

We begin with RDBMS-related issues (§4.1), then look at SQL-specific issues (§4.2), middleware-level challenges (§4.3), and finally analyze the management/administration of replicated databases (§4.4).

## 4.1 RDBMS-level Challenges

There are numerous challenges faced at the level of the database engine itself. The ones we have encountered most frequently are queries spanning multiple database instances, lack of flexibility in isolation levels, problems introduced by cluster heterogeneity, the handling of temporary tables, and suitable access control.

### 4.1.1 Multi-Database Queries
An RDBMS can manage multiple database instances (as created by CREATE DATABASE) and queries can span instances. Triggers, for example, are often used to perform reporting tasks and may update a different reporting database instance. Research efforts, however, focus primarily on replicating independent database instances [17] and have led to the concept of virtual databases. Virtualization of an entire RDBMS has not been addressed, so queries spanning multiple databases are usually not handled correctly by replication that works on a per-database-instance basis. Furthermore, RDBMSes generally lack mechanisms for taking a consistent snapshot of multiple databases.

Databases are sometimes also used as substitutes for schemas (as in CREATE SCHEMA). Some systems do not support the notion of schema at all (like MySQL), while for others, isolating data in different database instances is preferable over schemas for performance or security reasons.

Managing databases and schemas in replicated systems can sometimes be solved with engineering tweaks, but novel algorithms are still required when synchronizing multiple databases, in order to prevent synchronization of all databases according to a single total serializable order. Only cross-database accesses must be synchronized, but separating these from the overall workload requires the middleware to have a complete view of all database instance accesses.

### 4.1.2 Isolation Level
Database replication research has been largely addressing snapshot isolation (SI) and its variations (GSI, strong SI, weak SI, strong session SI, etc. [22]) in order to provide client applications with a behavior as close to 1-copy serializability as possible. SI is provided by Oracle (strongest isolation), DB2, PostgreSQL and, more recently, Microsoft SQL Server 2005. Database systems, such as Sybase or MySQL, do not provide SI. Nevertheless, the default setting in all DBMS is the weaker read-committed form, which most production applications use for performance reasons. Applications try to avoid transaction aborts with SI or deadlocks when using multi-version concurrency control (MVCC) at all costs. Enhancing current replication frameworks to support multiple isolation levels efficiently under a weaker isolation assumption is still an open area of research.

Related to isolation, the handling of request failures is different in the various implementations. The reasons why a request might fail range from malformed SQL to integrity constraint violations. PostgreSQL, for instance, aborts a transaction as soon as an error occurs, whereas MySQL continues the transaction until the client explicitly rolls back or aborts. To the best of our knowledge, no study has investigated error handling in replicated databases. This is a real problem in practice, especially when errors are combined with database updates that cannot be rolled back (DDL, auto-incremented keys, sequences, etc.).

### 4.1.3 Heterogeneous Clustering
*Hardware heterogeneity* is a fact of life in clusters of replicated databases. Heterogeneity, along with the ensuing unpredictability and performance disparity, inevitably occurs over time, because hardware components have limited lifetimes and replacing a piece of hardware by the same model 6 months later is difficult and not cost-effective. Heterogeneity issues can sometimes be addressed using dynamic load balancing techniques, such as LPRF [8].

Even when the cluster's hardware is homogeneous, the larger it is and the longer it has been in operation, the higher the variance in hardware and software performance, the higher the skew in data layout affecting disk throughput, workload asymmetries, etc. A few examples: a RAID controller with battery-backed write-back caches suddenly becomes 2x slower when the battery fails, and the OS rarely finds out; when a disk is replaced in a RAID-5, reconstruction severely impacts array performance; if a single strand in an Ethernet cable is crimped, throughput can drop from 1 Gbps to 100 Mbps. These anomalies are not addressed or evaluated by research focusing on load balancing strategies [4].

*Software heterogeneity* can be of 2 kinds: (1) It might be necessary to run multiple versions of the same database engine (more on this under software upgrades—see Section 4.4.3), or (2) aggregate data may be stored in two different database engines.

The first scenario usually happens during a migration phase, or when a legacy application that cannot be updated requires an older version of the database engine. As different versions of the database might require different drivers, either the application or the replication middleware have to ensure the proper driver is used in accordance with the accessed replica. Moreover, new functionality must be properly identified, so that queries using new features are only forwarded to capable replicas. Replication systems based on log shipping using binary formats have to handle all versions of storage formats and operate appropriate conversions.

The second scenario (aggregating data from multiple heterogeneous sources) is often seen when consolidating data from different departments of a company. Another common case occurs when companies are merged and databases have to be accessed from applications as a single data source. Replication adds complexity to these use cases, that already bring their own challenges. One option is for the application to use the smallest common denominator for all databases involved. Even pure ANSI SQL 92 compliant queries might not be able to execute similarly on all databases. Furthermore, some applications use middleware or object relational mappers that generate the queries on their behalf for a given database. In that case, the replicated database has to present itself as a data source of a specific kind (e.g., database engine A) and the replication middleware adapts automatically queries for database engines of other types. This adaptation can take the form of on-the-fly query rewriting, like in C-JDBC [8]. Also, Ganymed [29] scales a master database with different satellite databases that can possibly have a different schema and run different queries than the master.

### 4.1.4 Temporary Tables

Temporary tables are often used to simplify queries and store processed data between requests or transactions. The visibility of a temporary table varies between databases. In some cases, it is global, requiring a unique name, whereas in other implementations it is only visible to the current transaction or connection. Sybase, for instance, does not authorize the use of temporary tables within transactions. This forces the replication middleware to keep track of temporary tables so that connections stick to the same replica while using a given temporary table.

Temporary tables are not always persistent and are rarely backed up, even though they can persist across transaction boundaries. Management operations dealing with database backup/restore operations to bring new replicas online must make sure that no temporary tables are in use when a snapshot is taken, because this information cannot be made part of the snapshot.

The lack of conventions on temporary tables makes it difficult for the replication middleware to detect the true lifespan of a temporary table. Most applications do not explicitly delete temporary tables, but rather drop the connection, allowing the database to automatically free the corresponding resources on its own. Other implementations free temporary tables at commit time. Such drastic differences make it nearly impossible to implement a generally applicable middleware replication solution for temporary tables.

### 4.1.5 Access Control

Every connection starts with an authentication phase. Over time, databases have accommodated popular authentication mechanisms supporting a wide variety of access control methods. Middleware-based replication systems that intercept connections necessarily tamper with the database authentication mechanisms by hiding the original location of the client. However, it is necessary to capture client information, so that requests are replayed on behalf of the right user; as each user may have their own set of triggers, the same SQL statement might have a different impact, depending on which user is executing it.

The lack of user information standardization in the DBMS results in ad-hoc configurations and settings for each implementation. Despite the recent trend to store user data in the database information schema, access control information is often considered orthogonal to database content. This is a major problem when databases need to be cloned (even more so when it is a complete RDBMS with multiple database instances). Backup tools typically capture only data, without user-related information, raising issues when trying to clone a replica. Note that triggers and stored procedures are also rarely backed up (e.g., in ETL—Extraction, Transformation and Loading—tools, that focus mainly on data transformation without addressing user-related information).

## 4.2 SQL-level Challenges

In this section we discuss two main challenges induced by SQL semantics: stored procedures and large objects.

### 4.2.1 Stored Procedures

Stored procedures were initially introduced by Sybase and have been heavily used since then, with many legacy applications relying on stored procedures. The integration of Microsoft SQL Server 2005 with the .NET CLR has expanded the use of stored procedures [32] by allowing them to access thousands of pre-built classes and routines of the .NET Framework Base Class Library (BCL). Replication of stored procedures, however, raises several issues.

Statement replication can only broadcast calls to stored procedures, so stored procedure execution must be deterministic, to prevent cluster divergence. As there is no schema describing the behavior of a stored procedure, it is usually impossible to know which tables it accesses, thereby limiting concurrency control options for the middleware. Moreover, by replicating a stored procedure call, all the read queries will be executed by all nodes, resulting in no speedup and thus a waste of resources.

As stored procedures are often used to manipulate large amounts of data without transferring it to the client, performing writeset[2] extraction in such a context would be expensive in terms of resources, thus making it impractical in many cases.

Stored procedure replication is a domain that has been mostly overlooked by the research community. Even in a master/slave context, most databases have significant limitations on stored procedure replication and require the user to expand the stored procedure definition with ad-hoc extensions. Similar issues can be observed with user-defined functions.

### 4.2.2 Large Objects

Large objects, whether text (CLOB) or binary (BLOB), are implemented differently in different database engines (like many SQL data types). Object-relational DBMSes provide object identifiers (OIDs) and an API to retrieve an object's content.

If the replication middleware relays request and results, it must track resources properly to prevent the stream from remaining open indefinitely upon user program errors or failures. Moreover, certain drivers provide fake streaming APIs and require the application to have enough memory to hold the entire object in memory. Hence, multiple large objects, when streamed simultaneously, may quickly overwhelm the replication middleware.

### 4.2.3 Sequences

Database sequences, used to generate unique or auto-incremented keys, have only been standardized in SQL-2003 [12]. Even if, in most implementations, sequences can be retrieved as part of the database schemas, these objects are not persisted in the transactional log. This results in the need for workarounds to backup and restore sequences consistently with the other data.

---

[2] *Writeset*: the set of data $W$ updated by a transaction $T$, such that applying $W$ to a replica is equivalent to executing $T$ on it.

Additionally, sequences are non-transactional database objects, so they cannot be rolled back. Sequence numbers generated for a failed query or transaction are lost and generate "holes" in the sequence of numbers. Moreover, sequence semantics vary significantly among implementations; in most implementations, they bypass isolation mechanisms such as MVCC and are subject to subtle ordering problems.

## 4.3 Middleware-level Challenges

Middleware-based replication uses a middleware layer between the application and the database engines to implement replication, and there are multiple design choices for how to intercept client queries and implement replication across multiple nodes. We describe the most common alternatives with their pros and cons.

*4.3.1 Intercepting Queries*

Query interception needs may force driver changes on the application side as database protocols evolve over time. A new driver on the application side might offer new functionality, such as support for transparent failover or load balancing. Moreover, protocol version implementation may vary from one platform and language to another. For example, each MySQL JDBC, ODBC and Perl driver has its own bugs and ways to interpret the protocols. The Microsoft SQL Server JDBC driver and the FreeTDS open source implementation also exhibit different behavior, even though they are based on the exact same TDS protocol, originally designed by Sybase. Hence, it is difficult for the middleware to infer the application's intentions from the various implementations of a protocol in different drivers. In addition, some drivers exploit loopholes in the protocols to carry information for database extensions, such as geographic information services (GIS). This makes it even more difficult for the middleware to distinguish illegitimate input from undocumented extensions.

Updating drivers on the client side can be a real showstopper for sites with large clusters of application servers. If a customer has, e.g., 500 client machines accessing a cluster of 4 database server nodes, updating the driver is orders of magnitude more complex than upgrading the four nodes.

While JDBC and ODBC cover a large portion of database access methods for most recent applications, native APIs are still widely used by PHP and legacy applications. Supporting all APIs on all platforms quickly becomes unrealistic; for example, MySQL provides 14 main programming APIs for a database engine that is used on 16 different platforms (14 x 16 = 224 combinations).

*4.3.2 Statement vs. Transaction Replication*

Multi-master replication can be implemented either by multicasting every update statement (i.e., statement replication) or by capturing transaction writesets and propagating them after certification (i.e., transaction replication). Both approaches face significant challenges when put in production with real applications.

**Non-deterministic queries** are an important challenge: statement-based replication requires that the execution of an update statement produce the same result on each replica. However, SQL statements may legitimately produce different results on different replicas if they are not pre-processed before being issued.

Time-related macros such as '*now*' or '*current_timestamp*' are likely to produce a different result, even if the replicas are synchronized in time. Simple query rewriting techniques can circumvent the problem by replacing the macro with a hard-coded value that is common to all replicas. Of course, all replicas must still be time-synchronized and set in the same timezone, so that read queries provide consistent results.

Other macros, such as '*random*' or '*rand*', cannot always be replaced by a statically computed random number. Consider a statement like 'UPDATE t SET x=rand()'—a database engine would assign a different random value to each row of table t. Rewriting the query to hardcode a value like 'UPDATE t SET x=5' assigns the same value to each row, which was evidently not the programmer's intention. In this case, transaction replication would do the right thing, while statement replication would not.

Other queries may have non-deterministic results. For example, SELECT … LIMIT can create non-deterministic results in UPDATE statements. In 'UPDATE FOO SET KEYVALUE='x' WHERE ID IN (SELECT ID FROM FOO WHERE KEYVALUE IS NULL LIMIT 10)', the SELECT does not have an ORDER BY with a unique index. Therefore, broadcasting such a statement can cause each replica to update a different set of rows leading to divergence in the cluster.

**Writeset extraction** is usually implemented using triggers, to prevent database code modifications. This requires declaring additional triggers on every database table, as well as changing triggers every time the database schema is altered. This can be problematic both from an administrative as well as a performance standpoint when applications use temporary tables. If the application already uses triggers, writeset extraction through triggers might require an application rewrite. Materialized views also need special handling, to avoid duplicate writeset extraction by the triggers on the view and those on the underlying tables.

Writeset extraction does not capture changes like auto-incremented keys, sequence values, or environment variable updates. Queries altering such database structures change the replica they execute on and can contribute to cluster divergence. Moreover, most of these data structures cannot be rolled back (for instance, an auto-incremented key or sequence number incremented in a transaction is not decremented at rollback time).

Statement-based replication, at least, ensures that all these data structures are updated in the same order at all replicas. With transaction replication, if no coordination is done explicitly from the application, the cluster can end up in an endless effort to converge conflicting key values from different replicas.

**Locking and performance** are harder issues in statement-based replication. In particular, locking granularity is usually at the table level, as table information can be obtained through simple query parsing; however, this limits performance. Finer granularity (i.e., row level) would require re-implementing a large fraction of the database logic inside the middleware. Moreover, the middleware locking regime might not be compatible with the underlying database locking, leading to distributed deadlocks between the databases and the middleware.

*4.3.3 Failover*

Failover requires one or more failure detection mechanisms, so that everyone in the system can identify the same set of faulty components. Most problems are related to network timeouts (explained in Section 4.3.4). The state-of-the-art in failover has surprisingly not evolved much during the past decade. Even if failover does not require a system reconfiguration with automated reconnection mechanisms, in-flight sessions are still lost.

MySQL provides automatic reconnection inside its drivers, and application servers, like WebLogic [5], use multiple connection pools (multipools) for failover purposes. These techniques, or the one proposed in [22], offer session failover, but not failover of the transactional context. To the best of our knowledge, Sequoia [30] (the continuation of the C-JDBC project) is the only middleware that provides transparent failover without losing transactional context. Failover code is available in the middleware to handle a database failure, and additional code is available in the client driver to handle a middleware failure. Fully transparent failover requires consistently replicated state kept at all components, and is more easily achieved using statement-based rather than transaction-based replication. In the latter case, the transaction is only executed at a single replica; if the replica fails, the entire transaction has to be replayed at another replica, which cannot succeed without the cooperation of the application.

Even though a replicated database could handle internal component failures transparently to the clients, there is currently no API to pause, transfer and resume transaction contexts. Phoenix/COM+ [24] enhances the .NET runtime to serialize ODBC connection state in the database and allows COM-based applications to recover and failover transparently. Application server clusters typically operate on top of a replicated database; in the case when an application server replica fails, there is no way for the other replicas to retrieve the database connections of the failed replica and continue its transactions—even though the underlying database is capable of transparent failover. This is a manifestation of the more fundamental problem of failures being treated in isolation by each tier in a multi-tier architecture. Systems need to define user sessions and transaction contexts that cross tier boundaries, in order to treat failover issues globally and to ensure transparent failover throughout the system.

Connection pools are usually a major issue for failback. At failure time, all connections to a bad replica will be reassigned to another replica, or just removed from the pool. When the replica recovers from its failure, it requires the application to reconnect explicitly to that replica; this can only happen if the client connection pool recycles aggressively its connections, but this defeats the advantages of a connection pool. Most database APIs do not provide information on the endpoint of a database connection. Therefore, it is not possible for the connection pool to distinguish between connections to different replicas. In order to implement new load balancing and failover/failback strategies in connection pools, more contextual information on database connections is needed through standard database APIs.

### 4.3.4 Networking

Replicated databases are a distributed system, so they have to deal with network communication and related problems. As data loss is not acceptable during normal operation, it is necessary to have reliable communication channels. Reliable failure detectors are critical for failover and failback.

#### 4.3.4.1 Group Communication

A large body of research has been devoted to group communication protocols (a survey appears in [17]). Database replication requires reliable multicast with total order to ensure that each replica applies updates in the same order. Even though various optimizations have been developed, the group communication layer is an intrinsic scalability limit for such systems.

Group communication performance varies according to a large number of parameters [1], making configuration and tuning a real headache in practice. Even the developers of Spread, the most widely used group communication toolkit, admit that tuning UDP-based group communication is challenging even to a specialist. There is a subtle multi-dimensional tradeoff between ease of configuration (static vs. dynamic group membership), performance (UDP multicast vs. TCP performance in network switches, UDP multicast parallelism vs. TCP packet duplication, etc.), flow control (TCP flow control vs. credit-based flow control on UDP) and reliability (TCP in-kernel implementation with KeepAlive timeouts vs. UDP user-space error management implementations with tunable timeouts). Even though some issues can and must be addressed at the group communication level, cooperation with the replication middleware is key. For example, it is inefficient to perform state transfers when a new replica joins a cluster using group communication, because of the large amount of state to transfer.

Recent efforts have tried to extend multi-master replication to WAN environments [23]. The network latency and unreliability of long distance links are still making it impractical to have any reasonable production implementation of fast reliable multicast. Even though bandwidth availability is greatly improving, latency is unlikely to evolve dramatically on worldwide distances due to physical limitations. In practice, asynchronous replication is preferred over long distance links when replicating data between remote sites. Applications are usually partitioned and written using ad-hoc techniques to work around current technology limitations.

It is unlikely that group communication alone will be able to solve the database replication problem over WAN. 1-copy-serializability is unlikely to be successful in the WAN by extending existing LAN techniques. New data access models will have to be proposed to address the fundamental differences that one has to face when replicating in a WAN environment.

#### 4.3.4.2 TCP/IP Communication

Database drivers currently communicate with DBMSes through TCP connections, because TCP offers reliable communication with flow control that is efficiently implemented in the operating system kernel. However, TCP relies on timeouts to detect connection failures. Even though it is technically feasible to set up TCP timeouts on a per-connection basis, all drivers we know of rely on the default system-wide settings.

Upon a network failure, the TCP communication is blocked until the keep-alive timeout expires. This results in unacceptably long failure detection (ranging from 30 seconds to 2 hours, depending on the system defaults). Even though external heartbeat mechanisms are used to detect node failures, connections remain blocked on the client or server or both sides until the TCP timeouts expire. It is impractical to devise failover solutions for in-flight transactions over TCP, unless the underlying OS can be configured for each installation.

Altering operating system settings for the TCP KeepAlive value affects all applications running on that machine, and that is usually undesirable. A shorter TCP KeepAlive value generates false positives under heavy load by classifying slow connections as failed. Database drivers must either not rely on TCP for database communication, or have a built-in heartbeat mechanism for reliable and timely detection of connection failures.

### 4.3.4.3 Network Partitions

The issue of network partitions or "split brain" has been addressed by the research community mostly at a theoretical level; typically quorum-based solutions [19] are used. In practice, however, nodes often fail simultaneously (e.g., due to a rack-level power outage, or a network switch failure). If the remaining quorum does not constitute a majority, the system must shut down and make the customer unhappy.

The "CAP Principle" [15] states that a distributed storage system can favor any two of *C*onsistency, high *A*vailability, or resilience to *P*artitions. Systems such as search engines usually favor A and P over C, whereas a replicated database necessarily must favor C and A over P. The most common approach to handling network partitions is, therefore, to try and avoid them. If a network partition occurs, detecting it can be challenging, especially if the partition was due to a transient failure. When the system is partitioned, updating each partition independently leads to replica divergence. Some ETL and reconciliation tools do exist for fixing this [7], but the process remains largely manual; reconciliation policies are typically ad-hoc and application-dependent.

Partitions over WAN configurations usually require manual intervention of the human administrators at the various sites. If the network is indeed down, phone calls are usually used to diagnose the failure accurately and to coordinate a plan of action. The failover procedure usually has a wider scope than just the database replication system and typically involves DNS failover and other network-related reconfigurations.

## 4.4 System Management-level Challenges

Performing backups and adding/removing replicas are standard management operations for replicated databases. However, many customers desire at least 5 nines of availability for their database replication systems (99.999% availability, or at most 5 minutes of downtime per year), including all planned and unplanned downtime—this places tremendous pressure on the administrators. In this section we highlight some of the main challenges in managing replicated database systems: backup (§4.4.1), adding/removing replicas (§4.4.2), software upgrades (§4.4.3), routine maintenance (§4.4.4), and performance evaluation (§4.4.5).

### 4.4.1 Backup

Backup is part of normal database system operation, but is also fundamental in a replicated system, because backups are used to bring new replicas up-to-date. For most problematic backup and restore operations, databases can be taken offline (cold backup). ETL tools usually use database-specific extensions to access information such as user access rights, stored procedure or trigger definitions. Hot backup techniques exist, but they are still limited, because they only provide a read-consistent copy of the database, without handling active transactions. Database performance is typically degraded during backup. For example, in Oracle, when a database block is modified for the first time since the backup started, the entire block is written into the online redo logs. Under normal operation, only the changed bytes are written.

Since backup operations can take several hours, depending on how large the database is, it is important for hot backups and incremental backups to interplay with the replication middleware. It is unreasonable to expect applications that use large databases with high update rates to rely on cold backups with replication, since the backup time is not only the time it takes for the data to be dumped, but also the time needed to resynchronize the replica by reapplying all updates missed while doing the backup.

Thus, it is necessary for the replication middleware to collaborate with the replica and the backup tool, to make sure that the dumped data is consistent with respect to the entire cluster. This means that the middleware must be aware of exactly which transactions are contained in the dump and which ones must be replayed (or have their writesets applied), to properly resynchronize a backend. Replication middleware that supports partial replication affords a variety of optimizations for backup/restore.

### 4.4.2 Adding/Removing Replicas

Over time, replicas have to be removed from the system, usually for maintenance operations. If the replica is removed from the system due to a failure, a recovery operation is needed. Many systems, like MySQL cluster, require the entire cluster to be shut down and all replicas to be synchronized offline when adding new replicas. This implies long downtimes and unhappy customers.

Other solutions, like Emic Networks m/cluster, systematically use an active replica, bring it offline to transfer its state to the added replica, and then apply to both replicas the updates that occurred during transfer. This has the disadvantage of bringing the system down when only one replica is left in the system. Also, a node has to be taken offline when adding a new replica, which reduces performance during the operation. If the new replica is added with the intention of boosting performance, the operation has to be carefully planned, since overall system performance drops for the whole duration of the synchronization.

Sequoia [30] uses a recovery log that records all update statements executed by the system. When a node is removed from the cluster, a checkpoint is inserted, pointing to the last update statement executed by the removed node. When the node is re-added to the system, the recovery log is replayed from the checkpoint on. Offline nodes that have been properly checkpointed by the system can also be backed up. The resulting data dump can be used to initialize new replicas and resynchronize them from the recovery log, without having to use resources of active replicas.

Minimizing the cost of a cluster-wide checkpoint, while respecting transaction boundaries, is still an unsolved problem. Replaying the recovery log to resynchronize a replica requires the extraction of parallelism from the log to prevent reapplying updates serially, in which case a new replica may never catch up if the workload is update-heavy. Once a replica has replayed the entire recovery log, it is also necessary to enact a global barrier, to ensure that no in-flight request is missed by the newly-added replica. A large body of optimizations could be operated on replica synchronization, to minimize the resources and the time necessary to get to the online state.

Failures could often be recovered relatively easily. E.g., a replica might stop working because its log is full or its data partition ran out of space. However, the replication middleware has often no information on which transactions committed successfully prior to the failure; this information is only known to the database. As there is no standard API to query a database about the status of transactions, usually a full recovery has to be performed— in large production databases, this means hours of dump/restore and resynchronization. Fast resynchronization of failed nodes is as of yet still a hard problem.

Autonomic provisioning of database replicas [9] depends to a large extent on the system's ability to add and remove replicas. Being able to model and predict replica synchronization time and its associated resource cost is key to efficient autonomic middleware-based replicated databases. Determining the relevant metrics and exporting accurate resource usage predictions are challenges that need solutions before we can expect major breakthroughs in the area of autonomic replicated databases.

### 4.4.3 Software Upgrades

Software upgrades are part of planned maintenance operations. In a replicated database, there are three different types of components that can be upgraded: the engine itself, the replication middleware, and the drivers (database driver, middleware driver, or both, depending on the design).

*Database upgrade.* Database upgrades are usually relatively easy between minor releases, where a simple patch can be applied. Upgrades between major version numbers often require migration tools for both configuration and data files. If the replication middleware requires database modifications or extensions, it might not be possible to upgrade databases one by one without bringing the entire cluster down. A database upgrade while keeping the system online requires the replication middleware to support (at least temporarily) a heterogeneous cluster configuration, possibly using different driver versions for old and new database versions.

*Middleware upgrade.* As any software component, the replication middleware itself must be upgraded. If all components are replicated, it might be possible to upgrade them one by one, relying on standard failover techniques to handle the online upgrade. However, the protocols between replicated components must remain compatible, so that the old version of an upgraded component can still communicate with the newer version during the upgrade. This might require additional engineering, like one-time migration protocols, to allow upgrades between major versions of the replication middleware. If the replication middleware relies on a group communication library, upgrading the library requires protocol compatibility between versions.

*Driver upgrade.* Driver upgrades are rarely viewed as part of the database upgrade problem. However, it is quite common to have large web sites with tens or hundreds of application servers connecting to the same replicated database system. In such cases, upgrading the drivers is a much more complex issue than the database upgrade itself, which only affects a small number of machines and configurations.

### 4.4.4 Routine Maintenance

When running a production system, logs have to be purged, indexes have to be rebuilt, optimizer statistics need to be updated (vacuum-like operations), backups have to be made, etc. These operations have a significant impact on database performance both during and after their execution. Significant engineering efforts have gone into simplifying and automating database maintenance, but most of these efforts target centralized databases, leaving many open issues for replicated databases.

So far, there are no accepted "best practices" for performing maintenance on a replicated database system. What operations must be executed sequentially or in parallel? What is the impact on load balancing policies? Is it better to execute operations online or on an offline replica?

The vast majority of production systems have a monitoring infrastructure. Failures can be detected and reported by different sensors. It is not clear what interactions the database management framework should have with this global monitoring infrastructure. Whose responsibility is it to trigger an automatic replica repair operation when a failure is detected? A classification of replicated management operations is necessary in order to define the needed sensors, actuators, and to implement good management policies.

### 4.4.5 Performance Prediction

Database replication is often envisaged as a solution for performance issues. However, database replication usually only provides scalability, that is, if one adds resources proportionally to the load increase, the performance perceived per client will remain constant. Furthermore, the replication middleware itself imposes an overhead that often deteriorates query latency.

More insight is needed into the latency deterioration induced by moving from a single database to a replicated system, when the load could be handled without contention by the resources of a single database. We have observed that, when faced with workloads that have little parallelism, replicated databases usually perform poorly when load is low, because low latency is critical to the performance of sequential (non-parallel) queries. For example, a sequential batch update script will usually run much slower on a replicated database than on a single-instance database. OLTP-style sub-millisecond queries suffer the most from latency overheads imposed by the replication middleware, more so than heavyweight queries that take seconds or minutes to execute.

The lack of tools for accurately predicting the performance of a replicated database makes it difficult to properly size a system ("capacity planning") or to estimate the scalability limits of a system. As a result, database clusters tend to be small, between 2-4 replicas, rarely going up to 8 replicas. When it comes to OLTP databases, users feel safer to invest in fewer powerful machines than several less powerful machines. This helps limit complexity, as well as save on licensing and maintenance costs.

## 5. BRIDGING THE GAP

We now suggest a number of directions and areas for both academic research and industrial R&D, where innovative solutions could have a significant practical impact.

## 5.1 An Agenda for Academic Research

Tradeoffs for portability, upgradability, and performance vary across different designs for intercepting queries at the middleware level. Statement and transaction-based replication offer tradeoffs between performance, availability, and intrusiveness into the application or the RDBMS engine. Availability aspects, such as single points of failure cannot be overlooked. Failover and failback are tightly coupled with the usual networking issues and the black art of tuning timeouts. Performance is usually limited by group communication or writesets propagation. We see ample opportunities for optimization in systems operating under partial load or capacity, or at low consistency levels. A practical, deployment-worthy solution must address all these issues.

Research prototypes have mainly focused on performance, largely ignoring other system aspects. Every additional functionality, however, impacts the replication middleware's design, so the practicality of proposed concepts can only be assessed in a global context.

*Middleware design.* Database replication requires new abstractions in order to replicate more than one database instance at once. A RDBMS may host multiple database instances that appear as a single logical unit to the application or the customer. RDBMS replication poses new challenges for inter-database queries and cross-database management operations.

Partial replication is also a challenge: tables cannot be arbitrarily replicated, since queries might span over multiple tables and require distributed joins to perform select or update operations. Database backup is a complex distributed operation, since it might require multiple replicas to obtain a full consistent snapshot. Adding or removing partial replicas while still offering availability and service continuity is a completely open problem.

Stored procedure execution should be handled by the replication middleware. New algorithms are needed for optimizing the cluster-wide execution of stored procedures. If stored procedures were compiled in the middleware instead of the DBMS, queries and transactions could be better scheduled and balanced.

*Consistency.* SI and its variations attract substantial attention, as they improve performance over 1SR. Most probably, new optimizations or consistency models will be developed to address different needs. These new models, such as eventual consistency [34], could also require applications to be written differently, to better cooperate with the database and with new architectures, such as computing clouds. We ought to extract from past work the necessary interfaces and abstractions needed from both the replication middleware and the DBMS and to make these protocols pluggable in a replicated system. This would both encourage industry to provide standard APIs for replication and foster new research into other consistency models (e.g., targeting the very common read-committed transaction isolation level).

WAN environments impose both latency and availability constraints that are different from geographically centralized clusters. New consistency models for the WAN that are less strict than 1SR or SI, but stronger than fully asynchronous replication, require new protocols and probably programming paradigms.

*SPOF and Availability.* Production systems cannot tolerate any single points of failure, since service continuity is key. All management operations (e.g., backups or adding a replica) should be doable without service interruption. This requires fully transparent failover and failback that go beyond standard application/database boundaries. Availability might have to be thought of more globally, so that all failure detection mechanisms can synchronize to take proper coordinated actions. Connections and transactions should be addressed globally, so they can be transferred or failed over. Recovery procedures, distributed checkpointing, and replica state reconstruction are vast areas to investigate. Research on autonomic replicated databases ought to expand beyond performance, to tackle all aspects of availability.

Software upgrades are inevitable. All components including driver, middleware and database engine, must be upgradeable without service interruption. New solutions are required to allow such upgrades and to minimize the duration of these operations. A system with 5 nines of availability can be unavailable for no more than 5.26 minutes per year—this number marks the sole acceptable upper bound when evaluating new availability techniques. Similarly, metrics such as MTTF and MTTR should be considered when evaluating a design and/or prototype.

*Evaluation.* As database replication is about more than peak throughput, it is necessary to assess performance in the presence of failures, in degraded modes, as well as under low loads. Another area of performance that is not evaluated is the impact of management operations and faults on the system. New availability metrics should be defined, or combined with performance metrics, to better assess true overall performance.

To this end, researchers need new benchmarks that are not necessarily closed-loop systems, that could integrate fault injection or management operations. It would be interesting to have a wider variety of workloads, or to be able to capture workloads from existing applications. Even though it is possible to capture in various logs the execution of a workload, we know of no way yet to replay that exact same workload: the inherent parallelism in the original workload implies non-determinism in the execution order that is decided by the DBMS. Replaying a statistically equivalent workload is possible, but replaying the exact original workload at the same level of parallelism while still providing the same execution order requires instruction-level hardware simulation, which is still very expensive.

## 5.2 An Agenda for Industrial R&D

Database-agnostic middleware replication is challenging in practice, due to database implementation discrepancies. We have identified variations at the RDBMS level in transactional behavior, SQL semantics (e.g., temporary tables, stored procedures, large objects) and access control mechanisms. Despite efforts such as the GORDA project [16], the lack of standardization also affects management operations as well as recovery procedures. Hardware and software upgrades without service interruption lead to temporarily heterogeneous clustering and require innovative solutions.

*Integration.* Innovation in middleware-based replication with commercial databases has been limited to interactions using publicly available APIs. Workarounds, such as triggers for writeset extraction, have been implemented, but industry standards should be defined to better integrate with middleware replication. New mechanisms have to be developed to allow replication middleware to plug its own replacement for non-deterministic functions (e.g., time or random number functions).

*Transaction abstraction.* The notion of transaction as currently exposed by databases should be expanded to provide additional meta-information, such as readset and writeset, lock set, execution cost, etc. Having such information would allow for higher diagnosability at recovery time, would enable more efficient caches at the middleware level, and would improve decisions on which transactions to abort. Transactions are currently tightly coupled with driver connections—when the connection is lost, the transaction is lost; this precludes failover. It is currently not possible to pause a transaction, serialize and transfer transaction state to another connection, and resume a transaction.

*Management.* Backup/restore operations have to be improved to capture a consistent snapshot of a database without limiting themselves to data content. User information, access rights, views, triggers, and stored procedures must also be captured if a replica is to be properly cloned. The lack of standardization in this area thwarts further development of heterogeneous clustering with different database engines.

***Software upgrades.*** Finally, software upgrades are not only a problem for the database engine itself, but for all applications requiring a driver upgrade. New language runtimes, such as recent Java virtual machines, allow on-the-fly replacement of classes' implementations. This offers an infrastructure for dynamically upgrading drivers at the client side. The complexity of driver and database deployment could be considerably reduced by rethinking the driver lifecycle. Drivers used by client applications could be reduced to a minimum bootstrap, the database server providing the appropriate driver code at the first connection. Similar approaches could be used for database or middleware drivers.

# 6. CONCLUSION

In this paper, we reported challenges faced by middleware-based replication systems when deployed in production settings at real customers. We identified performance, availability and management issues that are insufficiently (or not at all) addressed by academic prototypes. Availability, in particular, poses several unsolved problems, from reliable failure detection to transparent failover/failback. Common management operations such as backups and hardware/software upgrades require the focused attention of the research community, to provide innovative solutions and optimizations. Database performance is not an issue that can be separated from availability and management issues.

We proposed four main themes to be investigated by academic research: replication middleware design, consistency, availability and evaluation. We also suggested new approaches to industrial R&D: improving the integration of databases with replication middleware, rethinking the way in which transactions are exposed to applications, standardizing management operations, and simplifying software upgrades.

# 7. ACKNOWLDEGEMENTS

We are grateful to our academic and industrial partners for their feedback and contributions and especially to Gustavo Alonso, Robert Hodges, Bettina Kemme, and Dheeraj Pandey for their detailed comments which helped improve this paper. This work was partially supported by NSF grants IIS-0133686, CCR-0205544, and IIS-0713409.